\documentclass[prb, aps, twocolumn]{revtex4}
\usepackage{graphicx}
\usepackage{bm}
\usepackage{amssymb}
\usepackage{amsmath}
\usepackage{amsfonts}
\usepackage{amsopn} 
\DeclareMathOperator{\trace}{tr}
\DeclareMathOperator{\Trace}{Tr}
\newcommand {\bra} [1] {\langle #1 |}
\newcommand {\ket} [1] {| #1 \rangle}
\newcommand {\pd} [2] {\frac{\partial #1}{\partial #2}}
\newcommand {\td} [2] {\frac{d #1}{d #2}}
\newcommand {\dbkt} [2] {\langle #1 | #2 \rangle}

\begin{document}

\title{Weak momentum scattering and the conductivity of graphene}
\author{Dimitrie Culcer and R. Winkler}
\affiliation{Advanced Photon Source, Argonne National Laboratory,
Argonne, IL 60439.} \affiliation{Northern Illinois University, De
Kalb, IL 60115.}

\begin{abstract}
  Electrical transport in graphene offers a fascinating parallel to
  spin transport in semiconductors including the spin-Hall effect.
  In the weak momentum scattering regime the steady-state density
  matrix contains two contributions, one linear in the carrier
  number density $n$ and characteristic scattering time $\tau$, the
  other independent of either. In this paper we take the Liouville
  equation as our starting point and demonstrate that these two
  contributions can be identified with pseudospin conservation and
  non-conservation respectively, and are connected in a non-trivial
  manner by scattering processes. The scattering term has a distinct
  form, which is peculiar to graphene and has important consequences
  in transport. The contribution linear in $\tau$ is analogous to
  the part of the spin density matrix which yields a steady state
  spin density, while the contribution independent of $\tau$, is
  analogous to the part of the spin density matrix which yields a
  steady state spin current. Unlike in systems with spin-orbit
  interactions, the $n$ and $\tau$-independent part of the
  conductivity is reinforced in the weak momentum scattering regime
  by scattering between the conserved and non-conserved pseudospin
  distributions.
\end{abstract}
\date{\today}
\maketitle

\section{Introduction}

The zero-gap semiconductor graphene, or two-dimensional carbon, is a
new material with a host of remarkable physical properties that
offers the possibility of all-carbon devices. The last three years
have seen a surge of experimental and theoretical interest in
graphene following its realization in the laboratory. \cite{gei07r,
net07r, kat07r} High-mobility graphene samples are nowadays reliably
manufactured, and the recent experimental success in applying a top
gate \cite{mar07} offers an increased handle on material properties.
Among the latest notable successes the ability to suspend graphene
experimentally \cite{sto08} is expected to help determine the value
of the dielectric constant $\epsilon$ unambiguously in this
material.

The linear spectrum characterizing the band structure of
single-layer graphene is reminiscent of dispersion relations in
relativistic physics, and the charge carriers in this material
behave like massless Dirac particles. Graphene has a honeycomb
lattice with two atoms per unit cell, and the Hamiltonian displays a
coupling between electron and hole states which gives rise to a
degree of freedom we refer to as the \textit{pseudospin}. It is
these facts that underlie its unusual features, which include the
vanishing of the density of states at the Dirac point, a
contribution to the conductivity independent of the carrier density
$n$ and scattering time $\tau$, a half-integer quantum Hall effect,
and the Klein paradox.\cite{gei07r, net07r, kat07r}

Recent experimental work includes the fabrication of epitaxial
graphene/graphene-oxide junction, \cite{wu07} measurement of
ultrafast carrier dynamics, \cite{daw07} shot noise measurements,
\cite{dan07, mar07b} determination of the performance limits of
graphene devices, \cite{chen07} observation of the Aharonov-Bohm
effect, \cite{rus07} observation of the quantum Hall effect near the
Dirac point, \cite{sto07} of a renormalization of the velocity due
to electron-phonon interaction, \cite{park07} and a thorough
experimental study of epitaxial graphene on SiC. \cite{var07}
Theoretical research on single-layer graphene has concentrated on,
among other matters, the effect of electron-electron interactions,
\cite{mis07} which are expected to be important due to weak
screening, on the question of whether graphene is a Fermi liquid,
\cite{tse07} and on the importance of localization around
impurities. \cite{per07} A number of theories have dealt with
scaling, \cite{schm07, bard07} impurity states, \cite{weh07} the
odd-integer quantum Hall effect, \cite{sheng07} the fractional
quantum Hall effect, \cite{tok06, chak06} polaritons, \cite{vaf06}
spin-orbit coupling \cite{bra06} and sum rules for the optical and
Hall conductivities. \cite{gus07} In addition to these, a large
number of theoretical predictions include the quantum spin-Hall
effect, \cite{kane05} spin-Hall conductance fluctuations,
\cite{qiao07} proximity-induced superconductivity, \cite{jiang07}
antiferromagnetism, \cite{bre07a} the spin-valve effect,
\cite{bre07b} peculiar focusing properties of graphene p-n
junctions, \cite{che07b} the use of graphene quantum dots as spin
qubits, \cite{trau07} Weiss oscillations, \cite{mat07} and a
zero-bias anomaly in the tunneling density of states. \cite{mari07}
Many other theories have sought to increase theoretical
understanding of graphene. \cite{her08, ludwig, sachdev, kat06,
aus07, bla07, nom06, zarb07, ber06, fer07, aba07, ada07a, ada07b,
yan07, zar07, mur06, bar07, zie06, zie07, fra86, ale06, che07a,
ost07, yao07, nov07, tru07, stau07, two06} Beyond the single-layer
form, graphene ribbons have been predicted to have spin polarized
edge states, \cite{aba06} while the band gap in bilayers has been
shown to be tunable by means of an electric field. \cite{cas06}

In the following we shall refer frequently to the weak momentum
scattering regime, characterized by $\varepsilon_F \gg \hbar/\tau$,
where $\tau $ is a characteristic momentum scattering time, and the
strong momentum scattering regime in which $\varepsilon_F \ll
\hbar/\tau$. Furthermore, it is conventional in the literature to
make the distinction between \emph{intrinsic} and \emph{extrinsic}
graphene. Intrinsic graphene refers to the specific case in which
the carrier doping density $n$ is zero, and the Fermi energy lies at
the Dirac point, $k = 0$. Intrinsic graphene is therefore by
definition in the strong momentum scattering regime. Extrinsic
graphene refers to the doped case and may be in either the weak or
the strong momentum scattering regime.

The presence of the additional degree of freedom contained in the
pseudospin causes the steady state for graphene in an electric field
to be qualitatively different from any other known material. In
particular, the presence of a contribution to the conductivity
independent of the carrier density remains a puzzling observation.
Such a contribution has been measured experimentally \cite{gei07r,
net07r, kat07r} and we emphasize that it was extracted from the
experimental data by taking the doping density to zero. The number
obtained is naturally characteristic of the strong momentum
scattering regime. It is typically referred to as the \emph{minimum}
conductivity of graphene. At the same time, theoretical research on
\emph{clean} graphene (no scattering) has found an additional
contribution to the conductivity independent of $n$ and $\tau$.
\cite{sachdev, stau07, fra86,ludwig, kat06, aus07, zie06, zie07,
two06} There appears to be some agreement that coherence between
electrons and holes lies at the heart of this particular
contribution to the conductivity. Since the Hamiltonian for carriers
in graphene is $\hbar v {\bm \sigma}\cdot{\bm k}$, where $v$ is a
constant and ${\bm \sigma}$ represents the pseudospin, the velocity
operator is simply $v {\bm \sigma}$, and it is evident that the
pseudospin will play a crucial role in transport. Nevertheless this
contribution to the conductivity was found in the ballistic regime,
whereas the contribution referred to as the \emph{minimum}
conductivity was measured experimentally in the strong momentum
scattering regime. We note that a series of enlightening papers have
focused on Boltzmann transport \cite{tru07}, on transport in the
strong momentum scattering limit \cite{nom06, her08} and transport
in extrinsic graphene \cite{ada07a, ada07b}. Our work differs from
these papers in that it is the first to recover both the ordinary
$\tau$-dependent ``Boltzmann'' conductivity and the $n$- and
$\tau$-independent conductivity, and to demonstrate the profound
relationship that exists between the two. In addition, our formalism
does not require us to work in the limit $n \rightarrow 0$ in order
to recover the $n$- and $\tau$-independent conductivity.

In order to arrive at a tractable equation for the density matrix it
is necessary to assume that the Fermi energy $\varepsilon_F \gg
\hbar/\tau$, which requires us to restrict our attention to the weak
momentum scattering regime. We assume low temperatures, where
scattering due to charged impurities is important and may be
dominant, and the fact that weakly doped graphene is a Fermi liquid
\cite{tse07} justifies omitting the effect of electron-electron
scattering. We believe that our approach sheds light on the manner
in which the part of the conductivity of graphene arises that is
independent of number density and $\tau$ and how it is related to
the ordinary, $\tau$-dependent (``Boltzmann'') conductivity. Indeed,
the theory presented in this work is tailored towards rendering
explicit the important role of the pseudospin in the dynamics of
carriers in graphene, with a focus on steady state processes,
providing an accurate and at the same time transparent approach. We
wish to emphasize that, although we determine the numerical value of
the conductivity, our principal aim is not to obtain a number but
rather to bring to light the underlying structure of the
steady-state density matrix in pseudospin space. Compared to the
Boltzmann picture, our formalism has coherence between electrons and
holes built in from the start and thus provides a clear physical
picture. Nevertheless weak localization effects\cite{nom06, her08}
are not taken into account in this work, as they are not expected to
be important in the weak momentum scattering regime.

We find the most important observation to be the fact that a
carrier's pseudospin is not conserved, because of electron-hole
coherence present in the Hamiltonian (including coherence induced by
the electric field). In the absence of intervalley scattering,
assumed in our work, a pseudospin eigenstate is an electron or a
hole, thus pseudospin non-conservation means a continually changing
combination of an electron and a hole. Therefore each carrier can be
thought of as a part which is either an electron or a hole, and a
part which is a continually changing mixture of an electron and a
hole. With this in mind, it makes sense to divide the pseudospin
density matrix into two linearly independent contributions,
corresponding to conserved pseudospin (the carrier is an electron or
a hole) and non-conserved precessing pseudospin (the carrier is a
continually changing mixture of an electron and a hole). We show in
a systematic fashion that the two contributions to the conductivity
of single-layer graphene are related to these linearly independent
components of the pseudospin density matrix. The $\tau$-dependent
contribution is a result of pseudospin conservation, while the $n$-
and $\tau$-independent conductivity stems from pseudospin
non-conservation. The two independent parts of the density matrix,
often referred to as the \emph{dissipative} and \emph{reactive}
parts, which are responsible for the two terms in the conductivity,
are connected in a non-intuitive way by scattering events. We find
that scattering gives rise to a term in the Liouville equation which
is peculiar to graphene and distinct from the usual scattering term
in other conductors. In particular, even an infinitesimal amount of
scattering produces a contribution to zeroth order in the scattering
potential which reinforces the $n$- and $\tau$-independent
contribution to the conductivity coming from the band structure.
This feature is one of the many intriguing analogies which exist
between transport in single-layer graphene and the generation of
steady-state spin densities and currents, including the spin-Hall
effect, in ordinary semiconductors. \cite{dim07, inoue, dim05} From
a different perspective, the two independent parts of the density
matrix are intimately related with the phenomenon of
\emph{zitterbewegung} which refers to a highly oscillatory component
in the motion of relativistic particles. \cite{sch30} Quite
generally, such an oscillatory component is always observed when the
electron states in two or more neighboring bands interfere.
\cite{win07} Of course, such an interference also lies at the heart
of (psuedo)spin precession so that the unique transport properties
of graphene can be considered to be a manifestation of
\emph{zitterbewegung} in a solid-state system. \cite{kat06}

The outline of the paper is as follows. In Sec.~II we derive a
kinetic equation for the density matrix specific to graphene, taking
the quantum Liouville equation as our starting point and focusing on
the form of the scattering term, which is different in graphene from
that in other materials even in the first Born approximation.
Following that, in Sec.~III we consider the dynamics of the
pseudospin. We divide the pseudospin density matrix into a part
representing conserved pseudospin and a part representing
non-conserved pseudospin, and demonstrate the way scattering affects
both of these parts and connects one to the other. In Sec.~IV we
determine the steady-state solution in the presence of an electric
field and the electrical conductivity. We show that the ordinary,
$\tau$-dependent conductivity can be traced to the conserved
pseudospin distribution, while the number density and
$\tau$-independent conductivity is associated with the non-conserved
pseudospin distribution. These arguments are further developed in
Sec.~V. We illustrate in Sec.~VI the remarkable similarities between
charge transport in graphene and spin transport in semiconductors
with strong spin-orbit interactions, including the way the vertex
correction to spin currents in semiconductors has an analog in
graphene that reinforces the $n$- and $\tau$-independent
conductivity in the weak momentum scattering regime. We conclude
with a brief summary in Sec.~VII.

\section{Time evolution of the density matrix}

The system is described by a density operator $\hat \rho$ which
obeys the quantum Liouville equation
\begin{equation}
\td{\hat\rho}{t} + \frac{i}{\hbar} \, [\hat{H} + \hat{H}^E
+ \hat U, \hat \rho] = 0,
\end{equation}
where $\hat{H}$ is the band Hamiltonian, $\hat{H}^E$ represents the
interaction with external fields, and $\hat{U}$ is the impurity
potential. We project the Liouville equation onto a set of
time-independent states of definite wave vector $\{ \ket{{\bm k}s}
\}$ that are not assumed to be eigenstates of $\hat{H}$. The matrix
elements of $\hat \rho$ in this basis are written as $\rho_{{\bm
k}{\bm k}'} \equiv \rho^{ss'}_{{\bm k}{\bm k}'} = \bra{{\bm k}s}
\hat\rho \ket{{\bm k}'s'}$, with corresponding notations for the
matrix elements of $\hat{H}$, $\hat{H}^E$ and $\hat U$, thus
$\rho_{{\bm k}{\bm k}'}$, $H_{{\bm k}{\bm k}'}$, $H^E_{{\bm k}{\bm
k}'}$, and $U_{{\bm k}{\bm k}'}$ are $2\times 2$ matrices in the
space spanned by the pseudospin $s$. We refer to $\rho_{{\bm k}{\bm
k}'}$ as the (pseudospin) density matrix. Matrix elements of the
band Hamiltonian $H_{{\bm k}{\bm k}'} = H_{{\bm k}} \, \delta_{{\bm
k}{\bm k}'}$ and $H^E_{{\bm k}{\bm k}'} = H^E_{{\bm k}} \,
\delta_{{\bm k}{\bm k}'}$ are diagonal in ${\bm k}$ but contain
off-diagonal terms in the pseudospin indices. Matrix elements of the
scattering potential $U_{{\bm k}{\bm k}'}$ are off-diagonal in ${\bm
k}$. (Matrix elements of the form $U_{{\bm k}{\bm k}}$ lead to a
redefinition of $H_{{\bm k}}$.) We assume elastic scattering and
work in the first Born approximation, in which $U_{{\bm k}{\bm
k}'}^{ss'} = U_{{\bm k}{\bm k}'}\delta_{ss'}$. Impurities are
assumed uncorrelated and the normalization is such that the
configurational average of $\bra{{\bm k}s}\hat U\ket{{\bm
k}'s'}\bra{{\bm k}'s'}\hat U\ket{{\bm k}s}$ is $(n_i |U_{{\bm k}{\bm
k}'}|^2 \delta_{ss'})/V$, where $n_i$ is the impurity density, $V$
the crystal volume and $U_{{\bm k}{\bm k}'}$ the matrix element of
the potential of a single impurity.

$\rho_{{\bm k}{\bm k}'}$ is divided into a part diagonal in ${\bm
k}$ and a part off-diagonal in ${\bm k}$, given by $\rho_{{\bm
k}{\bm k}'} = f_{{\bm k}} \, \delta_{{\bm k}{\bm k}'} + g_{{\bm
k}{\bm k}'}$, where, in $g_{{\bm k}{\bm k}'}$, it is understood that
${\bm k} \ne {\bm k}'$. We will be interested primarily in $f_{\bm
k}$ since most operators related with steady state processes are
diagonal in ${\bm k}$. The quantum Liouville equation is broken down
into
\begin{subequations}
\begin{eqnarray}
\label{eq:f} \td{f_{{\bm k}}}{t} + \frac{i}{\hbar} \,
[H_{{\bm k}}, f_{{\bm k}}] & = & - H^E_{\bm k} - \frac{i}{\hbar} \,
[\hat U, \hat g]_{{\bm k}{\bm k}} , \\ [1ex]
\label{eq:g} \td{g_{{\bm k}{\bm k}'}}{t} + \frac{i}{\hbar} \,
[\hat{H}, \hat g]_{{\bm k}{\bm k}'} & = & - \frac{i}{\hbar} \,
[\hat U, \hat f + \hat g]_{{\bm k}{\bm k}'}.
\end{eqnarray}
\end{subequations}
The solution to Eq.\ (\ref{eq:g}) to first order in $\hat{U}$ can be
written as
\begin{equation}\label{eq:gsol}
g_{{\bm k}{\bm k}'} = - \frac{i}{\hbar} \, \int_0^\infty dt'\,
e^{- i \hat H t'/\hbar} \left[\hat U, \hat f (t - t') \right]
e^{i \hat H t'/\hbar}|_{{\bm k}{\bm k}'},
\end{equation}
The assumption that $\varepsilon_F\tau/\hbar \gg 1$ allows us to
expand $\hat f(t - t')$ around $t$ and retain only $\hat f(t)$.
(Additional terms are of higher order in $\hat{U}$.) The equation
for $f_{{\bm k}}$ is
\begin{subequations}\label{eq:FermiJfk}
\begin{eqnarray}
\td{f_{{\bm k}}}{t} & + & \frac{i}{\hbar} \,
[H_{{\bm k}}, f_{{\bm k}}] + \hat J (f_{{\bm k}}) = - H^E_{\bm k}, \\ [1ex]
\label{eq:Jfk} \hat J (f_{{\bm k}}) & = & \frac{1}{\hbar^2} \!
\int_{0}^\infty \!\!\! dt'\, \left[\hat U, e^{- i \hat H t'/\hbar}
\left[\hat U, \hat f (t) \right] e^{ i \hat H t'/\hbar}
\right]_{{\bm k}{\bm k}}.
\end{eqnarray}
\end{subequations}
The integral in Eq.\ (\ref{eq:Jfk}) is performed by inserting a
regularizing factor $e^{- \eta t'}$ and letting $\eta \rightarrow 0$
in the end. For potentials $|U_{{\bm k}{\bm k}'}| \propto \openone$
which are scalars in pseudospin space this integral has the form
\begin{widetext}
\begin{equation}\label{eq:Jfinal}
\hat J(f_{\bm k}) = \frac{n_i}{\hbar^2} \lim_{\eta \rightarrow 0}
\int \frac{d^2k'}{(2\pi)^2} \, |U_{{\bm k}{\bm k}'}|^2
\int^{\infty}_0 dt'\, e^{- \eta t'}
\Bigl\{ e^{- i H_{{\bm k}'} t'/\hbar}\big(f_{\bm k} - f_{\bm k}' \big) \,
e^{i H_{{\bm k}} t'/\hbar} +  e^{- i H_{{\bm k}} t'/\hbar}
\big(f_{\bm k} - f_{\bm k}' \big) \, e^{i H_{{\bm k}'} t'/\hbar} \Bigr\}.
\end{equation}
\end{widetext}
Equation (\ref{eq:FermiJfk}) is a generalization of Fermi's golden
rule or, equivalently, a generalization of the first Born
approximation to systems, where the orbital motion is coupled with a
(pseudo)spin degree of freedom. \cite{ave02, dim07, ivc90}

\section{Pseudospin dynamics}

In the following we derive a scattering term specific to graphene
and an equation describing the time evolution of the pseudospin. The
band Hamiltonian for the carriers in each valley in single-layer
graphene at low doping densities is given by $H_{\bm k} = \hbar v \,
{\bm \sigma} \cdot {\bm k}$, where the constant $v$ is the Fermi
velocity and ${\bm \sigma}$ is the (two-dimensional) vector of Pauli
matrices in \emph{pseudospin} space. We emphasize that the
Hamiltonian does not depend on the true spin of particles, thus the
final result will contain a factor of 2 from the sum over the spin.
An additional factor of 2 must account for the valley degeneracy.
Consequently final results are multiplied by an overall factor of 4.

The Hamiltonian $H_{\bm k}$ is formally similar to the spin-orbit
interaction in spin-1/2 electron systems, \cite{win03} except that
the spin-orbit interaction is usually accompanied by a kinetic
energy term quadratic in $k$, which is typically much larger than
it, and has no analog in graphene. We wish to consider briefly this
aspect of $H_{\bm k}$ from the point of view of symmetry. The
Hamiltonian $H_{\bm k}$ transforms as a dipole in pseudospin space,
\cite{dim06} unlike the Hamiltonians of spin-1/2 electron systems,
which contain both a dipole (the spin-orbit interaction) and a
monopole (the scalar kinetic energy). In the absence of scattering
the equations of motion for the monopole and the dipole are
decoupled, \cite{dim06} but when scattering is present the time
evolution operator in the scattering term (\ref{eq:Jfinal}) in
general mixes the monopole and the dipole. In graphene the monopole
of the density matrix is equivalent to the scalar part and the
dipole is equivalent to the pseudospin part. Interestingly, because
the Hamiltonian for graphene only contains the dipole term,
scattering cannot mix the scalar and pseudospin parts of the density
matrix. We will see in the next paragraph how these symmetry
arguments become relevant.

\subsection{Scattering term}

Next we will evaluate the scattering term (\ref{eq:Jfinal}). For
this purpose, the density matrix of graphene is written as $f_{\bm
k} = n_{\bm k} + S_{\bm k} \equiv n_{\bm k} \openone + \frac{1}{2}\,
{\bm S}_{{\bm k}}\cdot {\bm \sigma}$ with a scalar part $n_{\bm k}$
and pseudospin part $S_{\bm k}$. We substitute $H_{\bm k}$ in the
time evolution operator in Eq.\ (\ref{eq:Jfinal}) and carry out the
time integration. In this way we obtain expressions for the action
of $\hat{J}$ on $n_{\bm k}$ and $S_{\bm k}$, and we find that, as
discussed above, this term does not mix $n_{\bm k}$ and $S_{\bm k}$.
The explicit expression for $\hat{J}(n_{\bm k})$ is not needed in
this work and will not be given. The action of the scattering term
on $S_{\bm k}$ is
\begin{equation}\label{eq:Jgph}
  \begin{array}[b]{@{}rl@{}}
    \hat J(S_{\bm k}) = & \displaystyle
    \frac{kn_i}{8\hbar \pi v} \int_{k'=k}
    d\theta' \, |U_{{\bm k}{\bm k}'}|^2 \big( {\bm S}_{{\bm k}} - {\bm
    S}_{{\bm k}'}\big) \\[3.2ex] & \displaystyle
    \times \big[ {\bm \sigma} \, (1 - \cos\gamma) +
    ({\bm \sigma}\cdot\hat{\bm k}) \, \hat{\bm k}' + \hat{\bm k} \,
    ({\bm \sigma}\cdot\hat{\bm k}') \big].
  \end{array}
\end{equation}
In the above $\theta'$ is the polar angle for the direction of ${\bm
k}'$ and $\gamma = \theta' - \theta$ is the angle between ${\bm k}'$
and ${\bm k}$. We see in Eq.\ (\ref{eq:Jgph}) that the scattering
term is qualitatively different from the scattering term in spin-1/2
electron systems,\cite{ave02} both in its angular dependence and in
not mixing the scalar and pseudospin parts of the density matrix.
Aside from the Born approximation, no further approximations were
made in deriving Eq.\ (\ref{eq:Jgph}). Yet we emphasize that it is
essential for this theory to assume weak momentum scattering since
there is no equivalent of the scalar kinetic energy term present in
semiconductors with spin-orbit interactions, and in order to derive
Eq.\ (\ref{eq:FermiJfk}) one must assume $\varepsilon_F\tau/\hbar =
v k_F \tau \gg 1$. The problem is characterized by two time scales,
the pseudospin precession frequency $\varepsilon_F/\hbar = v k_F$
and $\tau$, and in assuming $v k_F \tau \gg 1$ we were able to
truncate the scattering term at the leading order in the impurity
potential.

\subsection{Time evolution of the pseudospin}

We consider in more detail the equation for the pseudospin part of
the density matrix $S_{\bm k}$, in the general case in which a
nonzero source term exists on the RHS. Such a source term will be
present when an external field is acting on the system, and its form
can be derived straightforwardly from the quantum Liouville
equation. The specific case of an electric field will be discussed
in the following section. The kinetic equation for $S_{\bm k}$ in
the presence of a source $\Sigma_{\bm k}$ is
\begin{equation}\label{eq:Sk}
\td{S_{{\bm k}}}{t} + \frac{i}{\hbar} \, [H_{{\bm k}},
S_{{\bm k}}] + \hat J (S_{{\bm k}}) = \Sigma_{\bm k}.
\end{equation}
The structure of this equation is very important. In order to bring
it out we decompose $S_{\bm k}$ into two linearly independent parts,
$S_{\bm k} = S_{{\bm k} \|} + S_{{\bm k}\perp}$. $S_{{\bm k} \|}$
is, in matrix language, \emph{parallel} to the Hamiltonian $H_{\bm
k}$, while the remainder $S_{{\bm k}\perp}$ is \emph{orthogonal} to
$H_{\bm k}$. Because $[H_{{\bm k}}, S_{{\bm k} \|}] = 0$ the
parallel part $S_{{\bm k} \|}$ does not change in time under the
action of the time evolution operator $e^{iH_{\bm k}t/\hbar}$. In
other words $S_{{\bm k} \|}$ represents the fraction of the
pseudospin which is conserved, that is, carriers which are either
electrons or holes. Conversely, $S_{{\bm k}\perp}$ represents the
fraction of the pseudospin which is not conserved, i.e., it is
precessing. This fraction corresponds to carriers which are a
continually changing mixture of electrons and holes. The
decomposotion reflects the central importancee of electron-hole
coherence for the carrier dynamics in graphene. We remark that the
decomposition is fully equivalent to the decomposition of carrier
dynamics in relativistic quantum mechanics which contains a smooth
part and on oscillatory part known as \emph{zitterbewegung}.
\cite{sch30, win07} In that sense the physics discussed here
represents a direct manifestation of \emph{zitterbewegung} in
graphene. \cite{kat06}

Specifically, the following equations hold for $S_{{\bm k} \|}$ and
$S_{{\bm k}\perp}$
\begin{subequations}
\begin{eqnarray}
S_{{\bm k} \|} & = & \left[\frac{\trace (S_{\bm k} H_{\bm k})}
  {\trace (H_{\bm k}^2)}\right] H_{\bm k}, \\[1ex]
[S_{{\bm k} \|}, H_{\bm k}] & = & 0, \\ [1ex] \trace (S_{{\bm
k}\perp} H_{\bm k}) & = & 0,
\end{eqnarray}
\end{subequations}
where the symbol tr refers to a trace over the pseudospin indices
only. It is easily seen that
\begin{subequations}\label{eq:Sdecomp}
\begin{eqnarray} \label{eq:Sdecompa}
S_{{\bm k} \|} & = & {\textstyle\frac{1}{2}} \,
({\bm S}_{\bm k}\cdot\hat{\bm k}) ({\bm \sigma}\cdot\hat{\bm k})
\equiv {\textstyle\frac{1}{2}} \, s_{{\bm k} \|} \, \sigma_{{\bm
k} \|}, \\ [1ex]
S_{{\bm k}\perp} & = & {\textstyle\frac{1}{2}} \,
({\bm S}_{\bm k}\cdot\hat{\bm \theta}) ({\bm \sigma}\cdot\hat{\bm \theta})
\equiv {\textstyle\frac{1}{2}} \, s_{{\bm k}\perp} \, \sigma_{{\bm k}\perp},
\end{eqnarray}
\end{subequations}
where $\hat{\bm k}$ and $\hat{\bm \theta}$ are unit vectors along
the direction of ${\bm k}$ and perpendicular to ${\bm k}$,
respectively. We note that any matrix in pseudospin space can be
decomposed as in Eq.\ (\ref{eq:Sdecomp}). Therefore, in analogy with
this decomposition of the pseudospin part of the density matrix, we
also decompose the source term $\Sigma_{\bm k}$ along the same
principles. From Eq.\ (\ref{eq:Sk}) we can immediately see that the
equations describing the time evolution of $S_{{\bm k} \|}$ and
$S_{{\bm k}\perp}$ are
\begin{subequations}\label{eq:Spp}
\begin{eqnarray}
\td{S_{{\bm k} \|}}{t} + P_\| \hat J (S_{{\bm k}}) & = &
\Sigma_{{\bm k} \|}, \\ [0.5ex]
\td{S_{{\bm k}\perp}}{t} + \frac{i}{\hbar} \, [H_{{\bm k}}, S_{{\bm k}\perp}]
+ P_\perp \hat J (S_{{\bm k}}) & = & \Sigma_{{\bm k}\perp},
\end{eqnarray}
\end{subequations}
where $P_{\|/\perp} \hat J (S_{{\bm k}})$ indicates that the
scattering term acts on $S_{{\bm k}} = S_{{\bm k} \|} + S_{{\bm
k}\perp}$ and the resulting expression is projected
parallel/perpendicular to the Hamiltonian.

A solution for $S_{\bm k}$ can be found most straightforwardly by
expanding $S_{{\bm k} \|}$ and $S_{{\bm k}\perp}$ in the transition
rate $|U_{{\bm k}{\bm k}'}|^2$, in a manner analogous to that
adopted in determining the steady states of spin distributions in
systems with spin-orbit coupling. We found \cite{dim07} that in
steady-state problems the density matrix always contains a
correction $\propto \tau$ and is thus of order $-1$ in the
transition rate. This tells us that the expansion of $S_{\bm k}$
needs to start at order $-1$. Since we are working in the weak
momentum scattering limit we truncate this expansion at the next
highest order, which is order zero. The source term $\Sigma_{\bm k}$
does not have any dependence on the transition rate and is thus of
order zero. Equating terms of the same order in the transition rate
in Eq.\ (\ref{eq:Spp}) shows that the expansion of $S_{{\bm k} \|}$
must start at order $-1$, while the expansion of $S_{{\bm k}\perp}$
must start at order zero. We denote the order $-1$ in the transition
rate by a superscript $(-1)$ with the corresponding notation for
order zero. As a result Eqs.\ (\ref{eq:Spp}) in the weak momentum
scattering limit simplify to
\begin{subequations}\label{eq:Sppsimp}
\begin{eqnarray}\label{eq:Sppsimpa}
P_\|\, \hat J (S_{{\bm k} \|}^{(-1)}) & = &
\Sigma_{{\bm k} \|}, \\ [1ex] \label{eq:Sppsimpb}
\td{S_{{\bm k}\perp}^{(0)}}{t} + \frac{i}{\hbar} \, [H_{{\bm k}}, S_{{\bm
k}\perp}^{(0)}] & = & \Sigma_{{\bm k}\perp} - P_\perp\, \hat J
(S_{{\bm k} \|}) \\ [1ex] \label{eq:Sppsimpc}
P_\|\, \hat J (S_{{\bm k} \|}^{(0)}) & = &
- P_\| \, \hat J (S_{{\bm k}\perp}^{(0)}).
\end{eqnarray}
\end{subequations}
In Eq.\ (\ref{eq:Sppsimpa}) we have omitted the time-derivative of
$S_{{\bm k} \|}$ for the following reason. In both equations we are
looking for the steady state solution, and the equation for $S_{{\bm
k} \|}$ is most easily solved without the time derivative. The
equation for $S_{{\bm k}\perp}$ is most easily solved with the time
derivative explicitly taken into account, but the time dependence
drops out in the end.

Equation (\ref{eq:Sppsimp}) shows that, if the solution is required
only to order zero in the transition rate, the scattering term acts
only on $S_{{\bm k} \|}$, the part of the density matrix parallel to
the Hamiltonian. Physically, the fact that $S_{{\bm k} \|}$ starts
at a lower order in the transition rate than $S_{{\bm k}\perp}$
means that scattering processes are more effective at randomizing
the pseudospin than at scattering into pseudospin eigenstates.

Finally, we can simplify the scattering term by projecting Eq.\
(\ref{eq:Jgph}) onto and perpendicular to the Hamiltonian $H_{\bm
k}$. The projections that we will require in this work are
\begin{subequations}
  \begin{eqnarray}
    P_\| \hat J(S_{{\bm k} \|}) \! &\! = \! & \! \frac{kn_i}{8\hbar \pi v}\!\!
    \int\! d\theta' \, |U_{{\bm k}{\bm k}'}|^2\,
    (s_{{\bm k} \|} - s'_{{\bm k} \|})(1 + \cos\gamma)\,
    \sigma_{{\bm k} \|} , \nonumber\\[-0.5ex] \label{eq:Jspar} \\
    P_\perp \hat J(S_{{\bm k} \|}) \! &\! = \! & \! \frac{kn_i}{8\hbar \pi v}
    \!\! \int\! d\theta' \, |U_{{\bm k}{\bm k}'}|^2\,
    (s_{{\bm k} \|} - s'_{{\bm k} \|}) \sin\gamma \,
    \sigma_{{\bm k} \perp}, \nonumber\\[-0.5ex] \label{eq:Jsperp} \\
    P_\| \hat J(S_{{\bm k}\perp}) \! &\! = \! & \!
    \frac{kn_i}{8\hbar \pi v} \int\! d\theta' \, |U_{{\bm k}{\bm
    k}'}|^2\, \big(s_{{\bm k}\perp} + s_{{\bm k}'\perp}\big)
    \sin\gamma \sigma_{{\bm k} \|} .
    \nonumber \\
  \end{eqnarray}
\end{subequations}
We proceed to determine the concrete form of $\hat J(S_{{\bm k}
\|})$ for a screened Coulomb potential. The effect of screening in
graphene has been evaluated by Ando \cite{ando} among others, who
showed that $k_{TF} \propto k_F$, where $k_\mathrm{TF}$ is the
Thomas-Fermi wave vector. The explicit expression $k_{TF}$ will not
be reproduced here, it suffices to bear in mind that the ration
$k_{TF}/k_F$ is a constant. In two dimensions, the square of the
matrix element $U_{{\bm k}{\bm k}'}$ of a screened Coulomb potential
between plane waves is
\begin{subequations}
  \begin{eqnarray}
    |U_{{\bm k}{\bm k}'}|^2 & = & \frac{Z^2e^4}{\epsilon_0^2
    V_\mathrm{2D}^2}\, \frac{1}{4k^2\sin^2\frac{\gamma}{2} +
    1/L_s^2} \\ & \equiv &
    \frac{W}{\sin^2 (\gamma/2) + k_\mathrm{TF}^2 / k_F^2},
  \end{eqnarray}
\end{subequations}
where $Z = 1$ is the ionic charge, $L_s = k_F / (2k_\mathrm{TF}k)$
is the screening length. Substituting this into Eq.\
(\ref{eq:Jspar}) we obtain
\begin{subequations}
\label{eq:pJspar}
\begin{eqnarray}
P_\| \hat J(S_{{\bm k} \|}) & = & \frac{kn_iW}{4\hbar^2 \pi v} \int d\theta'
\, \zeta(\gamma) \, (s_{{\bm k} \|} - s'_{{\bm k} \|})\, \sigma_{{\bm k} \|}
\hspace{2.5em} \\[1ex]
\zeta(\gamma) & = & \frac{\cos^2(\gamma/2)}{\sin^2(\gamma/2)
+ k_\mathrm{TF}^2/ k_F^2}
\end{eqnarray}
\end{subequations}
with similar results for $P_\perp \hat J(S_{{\bm k} \|})$. In order
to evaluate this expression we expand $\zeta(\gamma)$ in a Fourier
series as $\zeta(\gamma) = \sum_m \zeta_m \, e^{im\gamma}$ and
remark that $\zeta_{-m} = \zeta_m$. In a similar way we expand
$s_{{\bm k} \|}$ as $s_{{\bm k} \|} = \sum_m s_{k\| m} \,
e^{im\theta}$. This gives for Eq.\ (\ref{eq:pJspar})
\begin{equation}\label{eq:Fourier}
P_\| \hat J(S_{{\bm k} \|}) = \frac{kn_iW}{2\hbar^2 v} \, \sigma_{{\bm k} \|}
\sum_m (\zeta_0 - \zeta_m) \, s_{k\| m} \, e^{i m \theta}.
\end{equation}
This is the furthest this equation can be simplified at this stage.
We will see below that in the steady state additional
simplifications emerge.

\section{Steady state solution}

In the following, we assume low fields $\bm E$ and look for a
solution to first order in $\bm E$. As shown in Appendix~A in the
presence of an electric field ${\bm E}$ the source term $\Sigma_{\bm
k}$ in Eq.\ (\ref{eq:Sk}) takes the form
\begin{equation}
  \Sigma_{\bm k} =
  \frac{e{\bm E}}{\hbar} \cdot \frac{\partial S_0}{\partial {\bm k}}.
\end{equation}
Here $S_0$ is the pseudospin part of the equilibrium density
matrix, i.e.,
\begin{equation}
S_0 = {\textstyle\frac{1}{2}}\, (f_{0+} - f_{0-}) \,\sigma_{{\bm k} \|} \, ,
\end{equation}
where the scalars $f_{0\pm} = f_0(\pm \hbar v k)$, with $f_0$ the
Fermi-Dirac distribution and $\pm \hbar v k $ the eigenenergies of
the graphene Hamiltonian $H_{{\bm k}}$, thus
\begin{equation}
  \label{eq:fermi}
  f_{0\pm} = \frac{1}{e^{\beta(\mp \hbar vk - \mu)} + 1},
\end{equation}
where $\mu$ is the chemical potential. The conserved and
non-conserved components of the source term are
\begin{subequations}
\begin{eqnarray}
\Sigma_{{\bm k} \|} & = &
\frac{e{\bm E}\cdot\hat{\bm k}}{2\hbar}
\, \left(\pd{f_{0+}}{k} - \pd{f_{0-}}{k}\right) \sigma_{{\bm k} \|}
\, , \label{eq:Sigmapar} \\[1ex]
\Sigma_{{\bm k}\perp} & = & \frac{e{\bm
E}\cdot \hat{\bm \theta}}{2 \hbar k} \, (f_{0+} - f_{0-})
\,\sigma_{{\bm k} \perp} \, .
\end{eqnarray}
\end{subequations}

\subsection{$S_{{\bm k} \|}$ to leading order in scattering}
\label{sec:Sparl}

Using Eqs.\ (\ref{eq:Sppsimpa}) and (\ref{eq:Fourier}) we can write
down the equation for $S_{{\bm k} \|}^{(-1)}$ and equate the
coefficients of $\sigma_{{\bm k} \|}$, reducing it to an equation
for $s_{{\bm k} \|}^{(-1)}$. It is important to note from Eq.\
(\ref{eq:Sigmapar}) that the coefficient of $\sigma_{{\bm k} \|}$
contains only the Fourier components $m = \pm 1$, and therefore the
sum in Eq.\ (\ref{eq:Fourier}) will only contain $\zeta_1$ (besides
$\zeta_0$). This means that we have for $s_{{\bm k} \|}^{(-1)}$
\begin{equation}
\frac{s_{{\bm k} \|}^{(-1)}}{\tau} = \frac{e{\bm E}\cdot\hat{\bm
k}}{2\hbar} \, \left(\pd{f_{0+}}{k} - \pd{f_{0-}}{k}\right).
\end{equation}
Here the scattering time $\tau$ is given by
\begin{subequations}
\label{eq:tau}
\begin{equation}
\frac{1}{\tau} = \frac{kn_iW}{2\hbar^2 v}
\, (\zeta_0 - \zeta_1) ,
\end{equation}
where the term in brackets can be expressed as
\begin{equation}
\zeta_0 - \zeta_1 = \left(\sqrt{1 + \frac{k_\mathrm{TF}^2}{k_F^2}} -
\frac{k_\mathrm{TF}}{k_F}\right)^2.
\end{equation}
\end{subequations}
Using Eq.\ (\ref{eq:Sdecompa}) this gives us $S_{{\bm k} \|}$
as
\begin{equation} \label{eq:Spar}
S_{{\bm k} \|}^{(-1)} = \frac{\tau\, e \,{\bm E}\cdot\hat{\bm
k}}{4\hbar} \left(\pd{f_{0+}}{k} - \pd{f_{0-}}{k}\right)
\sigma_{{\bm k} \|} \, .
\end{equation}

\subsection{$S_{{\bm k}\perp}$ to leading order in scattering}

We proceed to determine $S_{{\bm k} \perp}^{(0)}$. For this purpose
we require the term $P_\perp\, \hat J (S_{{\bm k} \|}^{(-1)})$ in
Eq.\ (\ref{eq:Sppsimpb}). Inserting Eq.\ (\ref{eq:Spar}) into Eq.\
(\ref{eq:Jsperp}) gives
\begin{equation}\label{eq:PJSpar}
P_\perp \hat J(S_{{\bm k} \|}^{(-1)}) = - \frac{e {\bm
E}\cdot\hat{\bm \theta}}{2 \hbar} \, \frac{\xi_0}{\zeta_0 - \zeta_1}
\left(\pd{f_{0+}}{k} - \pd{f_{0-}}{k}\right) \sigma_{{\bm k} \perp}
\, ,
\end{equation}
where the angular integral $\xi_0$ is given by
\begin{equation}
\xi_0 = \frac{1}{2\pi} \int d\theta' \, \frac{\sin^2
\frac{\gamma}{2} \cos^2\frac{\gamma}{2}} {\sin^2{\frac{\gamma}{2}} +
\frac{k_\mathrm{TF}^2}{k_F^2}}
= \frac{1}{2} \left(\sqrt{1 + \frac{k_\mathrm{TF}^2}{k_F^2}}
  - \frac{k_\mathrm{TF}}{k_F}\right)^2
\end{equation}
so that $\xi_0/(\zeta_0 - \zeta_1) = 1/2$ and Eq.\ (\ref{eq:PJSpar})
is indeed indpendent of the screening length. Now Eq.\
(\ref{eq:Sppsimpb}) for $S_{{\bm k}\perp}^{(0)}$ becomes
\begin{subequations}
\label{eq:Sperp}
\begin{equation}
\td{S_{{\bm k}\perp}^{(0)}}{t} + \frac{i}{\hbar} \,
[H_{{\bm k}}, S_{{\bm k}\perp}^{(0)}] = \frac{e{\bm E}\cdot
\hat{\bm \theta} \, \lambda(k)}{2\hbar} \, \sigma_{{\bm k} \perp} ,
\end{equation}
in which we have abbreviated the quantity
\begin{equation}
\lambda(k) = \frac{1}{k}\, (f_{0+} - f_{0-}) +
\frac{1}{2} \left(\pd{f_{0+}}{k} - \pd{f_{0-}}{k}\right).
\end{equation}
\end{subequations}
The solution of this equation is found most easily using the time
evolution operator $e^{iH_{\bm k}t/\hbar}$, and we allow the
electric field to have a small but finite frequency $\omega$, taking
the limit $\omega \rightarrow 0$ at the end,
\begin{equation}
\label{eq:Sperp3}
S_{{\bm k}\perp}^{(0)} = \frac{e{\bm
E}\cdot \hat{\bm \theta} \, \lambda(k)}{2\hbar} \,
\lim_{\eta, \omega \rightarrow 0} \frac{1}{2i(2vk - \omega - i\eta)}\,
\sigma_{{\bm k}\perp} .
\end{equation}
As discussed above, unlike $S_{{\bm k} \|}$ given by Eq.\
(\ref{eq:Spar}), the perpendicular part $S_{{\bm k}\perp}$ of the
spin density matrix is independent of the transition rate $|U_{{\bm
k}{\bm k}'}|^2$.

\subsection{$S_{{\bm k} \|}$ to zeroth order in scattering}

Finally, $S_{{\bm k} \|}^{(0)}$ is found from Eq.\ (\ref{eq:Sppsimpc})
\begin{equation}
P_\|\, \hat J (S_{{\bm k} \|}^{(0)}) = - P_\|
\, \hat J (S_{{\bm k}\perp}^{(0)}).
\end{equation}
Since $S_{{\bm k}\perp}^{(0)}$ is known, we need to take the
expression for $S_{{\bm k}\perp}^{(0)}$, act on it with the
scattering operator and project the resulting expression parallel to
$H$. This will then become the source term for $S_{{\bm k}
\|}^{(0)}$. The details of this process are given in Appendix B. The
result is
\begin{equation}
S_{{\bm k} \|}^{(0)} =
\frac{e{\bm E}\cdot\hat{\bm k} \, \lambda(k)}{2\hbar}
\lim_{\eta, \omega \rightarrow 0} \frac{1}{2i(2vk - \omega - i\eta)} \,
\sigma_{{\bm k} \|}.
\end{equation}
This term is very similar to $S_{{\bm k} \perp}^{(0)}$ and their
averages over directions in momentum space are the same.

\subsection{Electrical conductivity}

Using the velocity operator in single-layer graphene, given by $
{\bm v} = v \, {\bm \sigma}$, we can finally determine the
electrical current. The current operator depends on the pseudospin
and, following our reasoning so far, is decomposed into a parallel
part ($\|$) and a perpendicular part ($\perp$). The expectation
value of the current reads
\begin{equation}\label{eq:current2}
 j_x = - e\lim_{\eta \rightarrow 0} \int \frac{d^2k}{(2\pi)^2}
\left(v_{x{\bm k} \|} s_{{\bm k} \|} + v_{x{\bm k}\perp} s_{{\bm
k}\perp} \right).
\end{equation}
We convert the current tensor into the conductivity tensor ${\bm
\sigma}$ using Ohm's law ${\bm j} = {\bm \sigma} {\bm E}$. The
tensor ${\bm \sigma}$ is diagonal, with $\sigma_{xx} = \sigma_{yy}$
and $\sigma_{xy} = 0$. The contribution to the conductivity due to
$S_{{\bm k} \|}^{(-1)}$ (per valley and spin) is
\begin{equation}\label{eq:ord}
\sigma^\mathrm{ord}_{xx} = \frac{e^2}{4h} \, vk_F\tau
\end{equation}
$\sigma^\mathrm{ord}_{xx}$ behaves differently depending on the
nature of scatterers in the system. \cite{tru07} For long range
scatterers $\tau = (4\hbar^2\epsilon^2 v k_F)/(n_iZ^2e^4)$, with $Z$
the atomic number and $\epsilon$ the permittivity, thus
$\sigma^\mathrm{ord}_{xx}\propto (n/n_i)$. Short range scatterers
give $\tau = (2\hbar^2\epsilon^2v k_{TF}^2)/(n_iZ^2e^4k_F)$, but
since in graphene $k_{TF} \propto k_F$ as shown by Ando\cite{ando}
we still have the relationship $\sigma^\mathrm{ord}_{xx}\propto
(n/n_i)$.

The contribution due to $S_{{\bm k}\perp}^{(0)}$ requires a careful
evaluation of $\lim_{\eta, \omega \rightarrow 0} \int dk\, f_{0\pm}
/(2vk - \omega - i\eta)$. This is performed by replacing first
$1/(2vk - \omega - i\eta)$ by $i\pi \delta(2vk - \omega)$ while
keeping $\omega \ne 0$, then evaluating the integral using the
$\delta$-function. (Otherwise one can obtain a negative
conductivity.) We get
\begin{equation}
\sigma^{0\perp}_{xx} =  \lim_{\omega \rightarrow 0} \, \frac{\pi e^2}{8h}
\biggl[ \frac{1}{1 + e^{\beta(\mu + \hbar\omega/2)}}
     - \frac{1}{1 + e^{\beta(\mu - \hbar\omega/2)}} \biggr].
\end{equation}
Note that the limits $\omega \rightarrow 0$ and $\beta =
1/(k_\mathrm{B} T) \rightarrow \infty$ are not equivalent. Results
of a similar or identical magnitude have been found before \emph{in
the absence of disorder} and for $n=0$ only. \cite{fra86, zie06,
zie07, aus07, ludwig, kat06, two06} The contribution of
$S^{(0)}_{{\bm k} \|}$ to the conductivity is equal to (per valley
and spin) $\sigma^\mathrm{0\|}_{xx} = \sigma^\mathrm{0\perp}_{xx}$.
This result reflects the fact that $S_{{\bm k}\perp}^{(0)}$ and
$S_{{\bm k} \|}^{(0)}$ have the same angular average in momentum
space. Remarkably this holds for a screened Coulomb potential
regardless of the screening length.

\section{Discussion}

We would like to dwell on the role of the pseudospin in the
electrical conductivity of graphene. In our analysis the pseudospin
density matrix $S_{\bm k}$ is quite generally decomposed into a part
$S_{{\bm k} \|}$ representing conserved pseudospin (i.e., carriers
which are either electrons or holes) and a part $S_{{\bm k}\perp}$
representing non-conserved pseudospin (carriers which are a
continually changing mixture of electrons and holes). Pseudospin
non-conservation, which is crucial in determining the
scattering-independent contribution to the conductivity, occurs in
graphene due to the mixing of electron and hole states contained in
$H_{\bm k}$ (\emph{electron-hole coherence}). The steady state in
graphene therefore involves non-conservation of the pseudospin due
to the mixing of electron and hole states present in the
Hamiltonian. The derivation shown in the preceding section
demonstrates that the conserved pseudospin distribution gives rise
to the fraction $\sigma^\mathrm{ord}_{xx}$ of the conductivity given
by Eq.\ (\ref{eq:ord}). $\sigma^\mathrm{ord}_{xx}$ corresponds to
the ordinary electrical (``Boltzmann'') conductivity, which in the
steady state is proportional to the carrier number density $n$ and
the characteristic scattering time $\tau$. The non-conserved
pseudospin distribution gives rise to a second contribution to the
electrical conductivity, $\sigma^{0\perp}_{xx}$, which appears to be
independent of the carrier number density and scattering time.
However, our analysis shows that this is not the complete answer,
since scattering between $S_{{\bm k}\perp}$ and $S_{{\bm k} \|}$
produces an additional correction of order zero in the scattering
potential, $\sigma^{0\|}_{xx}$ which reinforces
$\sigma^{0\perp}_{xx}$. Scattering from the non-conserved pseudospin
distribution into the conserved pseudospin distribution is
represented by $P_\| \, \hat J (S_{{\bm k}\perp})$ in Eq.\
(\ref{eq:Sppsimpc}). In that sense scattering between the two
distributions has a constructive effect, and we emphasize that this
reinforcement holds for screened Coulomb scattering, regardless of
the screening length (that is, both short-ranged and long-ranged
impurity potentials.) Conversely, the additional correction
$\sigma_{xx}^{0\| b}$, which depends on the number density but not
on the impurity density, is also a result of scattering. We
emphasize that this reinforcement of the $n$- and $\tau$-independent
contribution to the electrical conductivity in graphene in the weak
momentum scattering limit was not found before and constitutes the
main result of our work. This work demonstrates the unity behind two
situations which until now appeared to be two different limits of
the problem of determining the electrical conductivity of graphene.
To date no approach has been put forward in which the Boltzmann and
$n$- and $\tau$-independent contributions to the conductivity are
treated on the same footing.

\section{Comparison with systems with spin-orbit interactions}

The derivation of the steady-state density matrix in graphene
presented in this work makes evident the many parallels which exist
between the establishment of electrical currents in graphene and the
establishment of steady-state spin densities and spin currents in
semiconductors with strong spin-orbit interactions. In a recent
paper \cite{dim07} we discussed the nature of the steady state in
systems with spin-orbit interactions, and showed that it is very
different from the steady state established in usual charge systems.
This difference is due to the presence of spin precession as a
result of spin-orbit coupling. In the steady state in spin-orbit
systems the spin density matrix is decomposed into a part
representing conserved (i.e., not precessing) spin and a part
representing precessing spin. The conserved spin distribution is
responsible for the establishment of a steady-state spin density,
which is proportional to the carrier number density and the
characteristic scattering time $\tau$. The precessing spin
distribution is responsible for steady state spin currents, which
are independent of the scattering time and in two dimensions appear
to be independent of the number density. Interestingly, the
correction equivalent to $S_{{\bm k} \|}^{(0)}$ vanishes in
spin-orbit systems. Our recent work \cite{dim07} demonstrated that
scattering from the conserved spin distribution into the precessing
spin distribution produces a correction to the precessing spin
distribution which in general acts to \emph{reduce} spin currents,
and in certain circumstances causes the spin current to be zero. In
that sense scattering between the two distributions also has a
destructive effect. Furthermore, our work showed that this
cancelation is due to the same physics that produces the vertex
correction to the spin conductivity in Green's functions approaches.
\cite{inoue}

Interestingly, the reinforcement of $\sigma^0_{xx}$ in graphene
happens in a way that is very similar to the vanishing of the spin
current mentioned above. In both cases the effect is caused by the
term $P_\| \, \hat J (S_{{\bm k}\perp}^{(0)})$ appearing in Eq.\
(\ref{eq:Sppsimpc}). Yet this term makes contributions of different
signs in the two systems. We remark that Eq.\ (\ref{eq:Sperp}) has
been solved by applying the time evolution operator. The product
$e^{-iH_{\bm k}t/\hbar} \sigma_{{\bm k}\perp} e^{iH_{\bm k}t/\hbar}$
is made up of two terms which have different dependencies on the
angle of the wave vector ${\bm k}$. The first term is proportional
to $\sigma_{{\bm k}\perp}$ and the second term is proportional to
$[H_{\bm k}, \sigma_{{\bm k}\perp}]$. The electrical current
operator in graphene is isotropic so that only the first term
contributes to the conductivity in this material. In contrast, the
spin current operator is proportional to ${\bm k}$. Thus only the
second term contributes to the spin current in spin-orbit coupled
systems. These results imply that the correction to the spin current
is brought about scattering out of $S_{{\bm k} \|}$ and into
$S_{{\bm k}\perp}$. On the other hand in graphene it is scattering
out of $S_{{\bm k}\perp}$ and into $S_{{\bm k} \|}$ that gives rise
to the correction to electrical conductivity. From the discussion
above we expect this reinforcement to appear as a result of the
vertex correction to the charge conductivity in graphene if a linear
response approach based on Green's functions is used.

In the field of spin transport a simple and elegant argument has
been formulated \cite{dim05} which explains why the spin current is
necessarily zero in certain systems. Briefly, the spin current is
proportional to the \emph{rate of change} of one of the spin
components, which must be zero in the steady state. A similar
argument \emph{cannot} be made for charge transport in graphene,
where the charge current is proportional to the pseudospin and does
not depend on the rate of change of any quantity in the steady
state. This is a reassuring observation: if an analogous argument
could be made it would imply that the electrical conductivity of
graphene vanishes identically, and this is evidently not the case.

Finally, we would like to point out that in spin-1/2 electron
systems with spin-orbit interactions, where the kinetic energy
greatly exceeds the spin-orbit splitting, it is customary to treat
the spin-orbit interaction as a perturbation. Yet it is interesting
to bear in mind that one expects results qualitatively similar to
those in graphene for a Rashba-type Hamiltonian with a very large
spin-orbit constant.

\section{Summary}

We have examined closely the nature of the steady state established
in graphene in the presence of an electric field. We have
demonstrated that the steady state in this material contains
important qualitative differences from the steady state in other
known conductors. The principal reason behind this difference is the
existence of a pseudospin degree of freedom, which is related to the
coupling between electrons and holes contained in the Hamiltonian.
In the weak momentum scattering regime there are two contributions
to the electrical conductivity in graphene, one linear in the
carrier number density and scattering time and one independent of
both. These contributions can be identified with pseudospin
conservation and non-conservation respectively, and are connected by
scattering processes. Scattering between the non-conserved and the
conserved pseudospin distributions doubles the contribution to the
conductivity independent of $n$ and $\tau$. Moreover, the
steady-state density matrix in graphene displays remarkable
similarities to the steady-state spin density matrix in systems with
spin-orbit interactions. The contribution linear in $n$ and $\tau$
has an analog in the part of the spin density matrix which yields a
steady state spin density, while the contribution independent of $n$
and $\tau$ is analogous to the part of the spin density matrix which
yields a steady state spin current. The reinforcement of the $n$-and
$\tau$-independent part of the conductivity is due to scattering
between the conserved and non-conserved pseudospin distributions.
This scattering also has an analogy in spin-1/2 electron systems
with spin-orbit interactions linear in ${\bm k}$, \cite{inoue,
dim05} except in those systems scattering between the conserved and
non-conserved spin distributions causes the spin-Hall current to
vanish.

We are very grateful to Allan MacDonald, Qian Niu, Branislav
Nikoli\'c, Horst St\"ormer, Philip Kim, Shaffique Adam, Bj\"orn
Trauzettel, Maxim Trushin, Di Xiao, Wang Yao, Wang Kong Tse, Handong
Chen, and Sun Wukong for stimulating discussions. The research at
Argonne National Laboratory was supported by the US Department of
Energy, Office of Science, Office of Basic Energy Sciences, under
Contract No. DE-AC02-06CH11357.

\appendix

\section{Steady state expectation values}

Our aim is to derive a kinetic equation for the density matrix in
the presence of an electric field ${\bm E}$. In order to evaluate
the electric field-induced source term in the kinetic equation we
start from the Liouville equation with the perturbing Hamiltonian
$H^E = e{\bm E}\cdot\hat{\bm r}$. The source term in the Liouville
equation is
\begin{equation}
\hat\Sigma = -\frac{ie{\bm E}}{\hbar}\cdot [\hat{\bm r},
\hat{\rho}_0],
\end{equation}
where $\hat{\rho}_0$ is the density operator in equilibrium, that
is, in the absence of the external field. We are interested in the
expectation value of an operator $\hat{O}$ (in our case the
electrical current), which is found by taking the trace of this
operator with the density matrix, and thus with the correction to
the density matrix due to the electric field. This correction
evidently depends on the source so that we consider first the trace
of the operator $\hat{O}$ with the source term due to the electric
field. This trace (denoted by Tr) is evaluated as follows
\begin{equation}
\Trace(\hat{O}\hat{\Sigma}) = -\frac{ie{\bm E}}{\hbar}\cdot \int
d^dr \, \trace\, O({\bm r}) \, [{\bm r}, \rho_0({\bm r})].
\end{equation}
We express $\rho_0({\bm r})$ in terms of its Fourier transform,
using the convention $\dbkt{\bm r}{\bm k} = e^{-i{\bm k}\cdot{\bm
r}}$, as
\begin{equation}
\rho_0({\bm r}) = \int \frac{d^dk}{(2\pi)^d}\int \frac{d^d
k'}{(2\pi)^d} \, e^{-i({\bm k} - {\bm k}')\cdot{\bm r}}\,
\rho_0({\bm k}, {\bm k}'),
\end{equation}
where $d=2$ for graphene. At this point we substitute the expression
for the spatially-inhomogeneous electric field and focus on the part
of the density matrix diagonal in wave vector
\begin{equation}
\Trace(\hat{O}\hat{\Sigma}) = \trace \int \frac{d^dk}{(2\pi)^d} \,
O({\bm k}) \, \frac{e{\bm E}}{\hbar}\cdot \pd{\rho_0}{{\bm k}}.
\end{equation}
This tells us that the source term in the kinetic equation is
therefore $(e{\bm E}/\hbar)\cdot (\partial\rho_0/\partial{\bm k})$.

\section{$S_{{\bm k} \|}$ to zeroth order in scattering}

We will discuss in this section the details of the contribution
$S_{{\bm k} \|}^{(0)}$, which is found from
\begin{equation}
P_\|\, \hat J (S_{{\bm k} \|}^{(0)}) = - P_\|
\, \hat J (S_{{\bm k}\perp}^{(0)}).
\end{equation}
As mentioned in the main text, $S_{{\bm k}\perp}^{(0)}$ is known. We
act on it with the scattering operator and project the resulting
expression parallel to $H$, yielding
\begin{equation}
\arraycolsep 0.3ex
\begin{array}{@{}rl@{}}
\displaystyle P_\| \hat J(S_{{\bm k}\perp}) = & \displaystyle
\frac{kn_i \lambda(k)}{8\hbar \pi v} \lim_{\eta, \omega \rightarrow 0}
\frac{1}{2i(2vk - \omega - i\eta)} \\ [3ex]
& \displaystyle \times \frac{e{\bm E}}{2\hbar}\cdot\int\! d\theta' \,
|U_{{\bm k}{\bm k}'}|^2\,
\big(\hat{\bm \theta} + \hat{\bm \theta}'\big) \sin\gamma
\, \sigma_{{\bm k} \|} .
\end{array}
\end{equation}
The term $\propto \hat{\bm \theta}$ vanishes in the angular
integration, while the remaining term gives
\begin{equation}
P_\| \hat J(S_{{\bm k}\perp}) = - \frac{e{\bm E}\cdot\hat{\bm k} \,
\lambda(k)}{2\hbar \tau}
\lim_{\eta, \omega \rightarrow 0} \frac{1}{2i(2vk - \omega - i\eta)} \,
\sigma_{{\bm k} \|}.
\end{equation}
This will then act as the source term for $S_{{\bm
k} \|}^{(0)}$, and the equation is solved in the same way as
in Sec.~\ref{sec:Sparl}
\begin{equation}
S_{{\bm k} \|}^{(0)} =
\frac{e{\bm E}\cdot\hat{\bm k} \, \lambda(k)}{2\hbar}
\lim_{\eta, \omega \rightarrow 0} \frac{1}{2i(2vk - \omega - i\eta)} \,
\sigma_{{\bm k} \|}.
\end{equation}


\begin{thebibliography}{10}
 \bibitem{gei07r} A.~K. Geim and K.~S. Novoselov, Nature Materials
  \textbf{6}, 183 (2007).

 \bibitem{net07r} A.~H. Castro Neto, F.~Guinea, N.~M.~R. Peres,
  K.~S.~Novoselov, and A.~K. Geim, arXiv:0709.1163, to appear in
  Rev.~Mod.~Phys.

 \bibitem{kat07r} M.~I.~Katsnelson and K.~S.~Novoselov, Solid State
  Commun. \textbf{143}, 3 (2007).

 \bibitem{mar07} J.~R.~Williams, L.~DiCarlo, and C.~M. Marcus,
  Science \textbf{317}, 638 (2007).

 \bibitem{sto08} K.~I.~Bolotin, K.~J.~Sikes, Z.~Jiang, G.~Fudenberg,
  J.~Hone, P.~Kim, and H.~L.~St\"ormer, Solid State Communications
  \textbf{146}, 351-355 (2008).

 \bibitem{wu07} X.~Wu, M.~Sprinkle, X.~Li, F.~Ming, C.~Berger, and
  W.~A. de~Heer, Phys.~Rev.~Lett. \textbf{101}, 026801 (2008).

 \bibitem{daw07} J.~M.~Dawlaty, S.~Shivaraman, M.~Chandrashekhar,
  F.~Rana, and M.~G.~Spencer, Appl.~Phys.~Lett. \textbf{92}, 042116
  (2008).

 \bibitem{dan07} R. Danneau, F. Wu, M.~F. Craciun, S. Russo, M.~Y.
  Tomi, J. Salmilehto, A.~F. Morpurgo, and P.~J. Hakonen,
  Phys.~Rev.~Lett. \textbf{100}, 196802 (2008).

 \bibitem{mar07b} L. DiCarlo, J. R. Williams, Y. Zhang, D. T.
  McClure, and C. M. Marcus, Phys.~Rev.~Lett. \textbf{100}, 156801
  (2008).

 \bibitem{chen07} J. H. Chen, C. Jang, S. Xiao, M. Ishigami, and M.
 S. Fuhrer, Nature Nanotechnology 3, 206 - 209 (2008).

 \bibitem{rus07} S. Russo, J. B. Oostinga, D. Wehenkel, H. B.
 Heersche, S. S. Sobhani, L. M.~K. Vandersypen, and A. F. Morpurgo,
 Phys. Rev. B \textbf{77}, 085413 (2008).

 \bibitem{sto07} Z.~Jiang, Y.~Zhang, H.~L. Stormer, and P.~Kim,
 Phys.~Rev.~Lett. \textbf{99}, 106802 (2007).

 \bibitem{park07} C.-H.~Park, F.~Giustino, M.~L.~Cohen, and S.~G.
 Louie, Phys.~Rev.~Lett. \textbf{99}, 086804 (2007).

 \bibitem{var07} F.~Varchon, R.~Feng, J.~Hass, X.~Li, B.~Ngoc~Nguyen,
 C.~Naud, P.~Mallet, J.-Y.~Veuillen, C.~Berger, E.~H. Conrad, and
 L.~Magaud, Phys.~Rev.~Lett. \textbf{99}, 126805 (2007).

 \bibitem{mis07} E.~G.~Mishchenko, Phys.~Rev.~Lett. \textbf{98},
 216801 (2007).

 \bibitem{tse07} S.~Das Sarma, E.~H.~Hwang, and W.-K.~Tse,
 Phys.~Rev.~B \textbf{75}, 121406 (2007).

 \bibitem{per07} V.~M.~Pereira, J.~Nilsson, and A.~H.~Castro~Neto,
 Phys.~Rev.~Lett. \textbf{99}, 166802 (2007); V.~M.~Pereira,
 F.~Guinea, J.~M.~B.~Lopes~dos~Santos, N.~M.~R.~Peres, and
 A.~H.~Castro~Neto, Phys.~Rev.~Lett. \textbf{96}, 036801 (2006).

 \bibitem{schm07} D.~E.~Sheehy and J.~Schmalian, Phys.~Rev.~Lett.
 \textbf{99}, 226803 (2007).

 \bibitem{bard07} J. H. Bardarson, J. Tworzydo, P. W. Brouwer, and C.
 W. J. Beenakker Phys. Rev. Lett. 99, 106801 (2007).

 \bibitem{weh07} T.~O.~Wehling, A.~V.~Balatsky, M.~I.~Katsnelson,
 A.~I. Lichtenstein, K.~Scharnberg, and R.~Wiesendanger, Phys.~Rev.~B
 \textbf{75}, 125425 (2007).

 \bibitem{sheng07} L.~Sheng, D.~N.~Sheng, F.~D.~M.~Haldane, and
 L.~Balents, Phys.~Rev.~Lett. \textbf{99}, 196802 (2007).

 \bibitem{tok06} C.~Toke, P.~E.~Lammert, J.~K.~Jain, and
 V.~H.~Crespi, Phys.~Rev.~B \textbf{74}, 235417 (2006).

 \bibitem{chak06} V.~M.~Apalkov and T.~Chakraborty, Phys.~Rev.~Lett.
 \textbf{97}, 126801 (2006).

 \bibitem{vaf06} O.~Vafek, Phys.~Rev.~Lett. \textbf{97}, 266406
 (2006).

 \bibitem{bra06} D.~Huertas-Hernando, F.~Guinea, and A.~Brataas,
 Phys.~Rev.~B \textbf{74}, 155426 (2006).

 \bibitem{gus07} V.~P.~Gusynin, S.~G.~Sharapov, and J.~P.~Carbotte,
 Phys.~Rev.~B \textbf{75}, 165407 (2007).

 \bibitem{kane05} C.~L.~Kane and E.~J.~Mele, Phys.~Rev.~Lett.
 \textbf{95}, 226801 (2005).

 \bibitem{qiao07} Z.~H.~Qiao, J.~Wang, Y.~D.~Wei, and H.~Guo, Phys.
 Rev. Lett. \textbf{101}, 016804 (2008).

 \bibitem{jiang07} Y. Jiang, D.-X. Yao, E. W. Carlson, H.-D. Chen,
 and J. Hu, Phys.~Rev.~B \textbf{77}, 235420 (2008).

 \bibitem{bre07a} L. Brey, H. A. Fertig, S. Das Sarma, Phys. Rev.
 Lett. \textbf{99}, 116802 (2007).

 \bibitem{bre07b} L.~Brey and H.~A.~Fertig, Phys.~Rev.~B \textbf{76},
 205435 (2007).

 \bibitem{che07b} V.~V.~Cheianov, V.~Fal'ko, and B.~L.~Altshuler,
 Science \textbf{315}, 1252 (2007).

 \bibitem{trau07} B.~Trauzettel, D.~V.~Bulaev, D.~Loss, and
 G.~Burkard, Nature~Phys. \textbf{3}, 192 (2007).

 \bibitem{mat07} A.~Matulis and F.~M.~Peeters, Phys.~Rev.~B
 \textbf{75}, 125429 (2007).

 \bibitem{mari07} E.~Mariani, L.~I.~Glazman, A.~Kamenev, and
 F.~von~Oppen, Phys.~Rev.~B \textbf{76}, 165402 (2007).

 \bibitem{sachdev} L.~Fritz, J.~Schmalian, M.~Mueller, and
 S.~Sachdev, Phys.~ Rev.~B \textbf{78}, 085416 (2008).

 \bibitem{stau07} T.~Stauber, N.~M.~R.~Peres, and F.~Guinea,
 Phys.~Rev.~B \textbf{76}, 205423 (2007).

 \bibitem{fra86} E.~Fradkin, Phys.~Rev.B \textbf{33}, 3263 (1986).

 \bibitem{zie06} K.~Ziegler, Phys.~Rev.~Lett. \textbf{97}, 266802
 (2006).

 \bibitem{zie07} K.~Ziegler, Phys.~Rev.~B \textbf{75}, 233407 (2007).

 \bibitem{aus07} M.~Auslender and M.~I.~Katsnelson, Phys.~Rev.~B
 \textbf{76}, 235425 (2007).

 \bibitem{ludwig} A.~W.~W.~Ludwig, M.~P.~A.~Fisher, R.~Shankar, and
 G.~Grinstein, Phys.~Rev.~B \textbf{50}, 7526 (1994).

 \bibitem{kat06} M.~I.~Katsnelson, Eur.~Phys.~J.~B \textbf{51}, 157
 (2006).

 \bibitem{two06} J.~Tworzydlo \textit{et al}, Phys.~Rev.~Lett.
 \textbf{96}, 246802 (2006).

 \bibitem{tru07} M.~Trushin and J.~Schliemann, Phys.~Rev.~Lett.
 \textbf{99}, 216602 (2007).

 \bibitem{nom06} K.~Nomura and A.~H.~MacDonald, Phys.~Rev.~Lett.
 \textbf{98}, 076602 (2007).

 \bibitem{her08} I.~Herbut, V.~Juricic, and O.~Vafek,
 Phys.~Rev.~Lett. \textbf{100}, 046403 (2008).

 \bibitem{ada07a} S.~Adam, E.~H.~Hwang, V.~M.~Galitski, and
 S.~Das~Sarma, Proc.~Natl.~Acad.~Sci.~USA \textbf{104}, 18392 (2007).

 \bibitem{ada07b} E.~H.~Hwang, S.~Adam, and S.~Das~Sarma,
 Phys.~Rev.~Lett. \textbf{98}, 186806 (2007).

 \bibitem{bla07} Ya.~M.~Blanter and I.~Martin, Phys.~Rev.~B
 \textbf{76}, 155433 (2007).

 \bibitem{zarb07} L.~P.~Zarbo and B.~K.~Nikoli\'{c}, Europhys.~Lett.
 \textbf{80}, 47001 (2007).

 \bibitem{ber06} B.~A.~Bernevig, T.~L.~Hughes, H.-D.~Chen, C.~Wu,
 S.-C.~Zhang, Int.~Jour.~Mod.~Phys.~B, \textbf{20}, 3257 (2006).

 \bibitem{fer07} J.~Fernandez-Rossier, J.~J.~Palacios, and L.~Brey,
 Phys.~Rev.~B \textbf{75}, 205441 (2007).

 \bibitem{aba07} D.~A.~Abanin, P.~A.~Lee, and L.~S.~Levitov, Solid
 State Comm. \textbf{143}, 77 (2007).

 \bibitem{yan07} X.-Z.~Yan, Y.~Romiah, and C.~S.~Ting, Phys.~Rev.~B
 \textbf{77}, 125409 (2008).

 \bibitem{zar07} M.~Zarea and N.~Sandler, Phys.~Rev.~Lett.
 \textbf{99}, 256804 (2007)

 \bibitem{mur06} K.~Sasaki, S.~Murakami, and R.~Saito,
 Appl.~Phys.~Lett. \textbf{88}, 113110 (2006).

 \bibitem{bar07} Y.~Barlas, T.~Pereg-Barnea, M.~Polini, R.~Asgari,
 and A.~H.~MacDonald, Phys.~Rev.~Lett \textbf{98}, 236601(2007).

 \bibitem{ale06} I.~L.~Aleiner and K.~B.~Efetov, Phys.~Rev.~Lett.
 \textbf{97}, 236801 (2006).

 \bibitem{che07a} V.~V.~Cheianov, V.~I.~Fal'ko, and B.~L.~Altshuler,
 and I.~L.~Aleiner, Phys.~Rev.~Lett. \textbf{99}, 176801 (2007).

 \bibitem{ost07} P.~M.~Ostrovsky, I.~V.~Gornyi, A.~D.~Mirlin,
 Phys.~Rev.~Lett. \textbf{98}, 256801 (2007).

 \bibitem{yao07} W.~Yao, D.~Xiao, and Q.~Niu, Phys.~Rev.~B
 \textbf{77}, 235406 (2008).

 \bibitem{nov07} D.~S.~Novikov, Phys.~Rev.~Lett. \textbf{99}, 056802
 (2007).

 \bibitem{aba06} D.~A.~Abanin, P.~A.~Lee, and L.~S.~Levitov,
 Phys.~Rev.~Lett. \textbf{96}, 176803 (2006).

 \bibitem{cas06} E.~V.~Castro, K.~S.~Novoselov, S.~V.~Morozov,
 N.~M.~R.~Peres, J.~M.~B.~Lopes~dos~Santos, J.~Nilsson, F.~Guinea,
 A.~K. Geim, A.~H.~Castro~Neto, Phys.~Rev.~Lett. \textbf{99}, 216802
 (2007).

 \bibitem{win03} R.~Winkler, {\it Spin-Orbit Coupling Effects in
 Two-Dimensional Electron and Hole Systems} (Springer, Berlin, 2003).

 \bibitem{dim06} D.~Culcer, C.~Lechner, and R.~Winkler,
 Phys.~Rev.~Lett. \textbf{97}, 106601 (2006).

 \bibitem{ave02} N.~S.~Averkiev, L.~E.~Golub, and M.~Willander,
 J.~Phys.:~Cond.~Mat. \textbf{14}, R271 (2002).

 \bibitem{dim07} D.~Culcer and R.~Winkler, Phys.~Rev.~B. \textbf{76},
 245322 (2007).

 \bibitem{ivc90} E.~L.~Ivchenko, Y.~B. Lyanda Geller, and G.~E.
 Pikus, Sov.~Phys.~JETP \textbf{71}, 551 (1990).

 \bibitem{ando} T.~Ando, J.~Phys.~Soc.~Jpn. \textbf{75}, 074716
 (2006).

 \bibitem{inoue} J.~Inoue, G.~E.~W.~Bauer, and L.~W.~Molenkamp, Phys.
 Rev. B \textbf{70}, 041303 (2004).

 \bibitem{dim05} O.~V.~Dimitrova, Phys. Rev. B \textbf{71}, 245327
 (2005).

 \bibitem{sch30} E. Schr\"odinger, Sitzungsber. Preuss. Akad. Wiss.,
 Phys. Math. Kl. \textbf{24}, 418 (1930).

 \bibitem{win07} R.~Winkler, U. Z\"ulicke, and J. Bolte,
  Phys.~Rev.~B \textbf{75}, 205314 (2007).

\end{thebibliography}
\end{document}